\documentstyle[aps,preprint,tighten]{revtex}
\begin{document}
\title{Kinetics of Bose-Einstein Condensation in a Trap}
\author{C.W.~Gardiner$^{1}$, P.~Zoller$^2$, R.J.~Ballagh$^{3}$ and 
M.J.~Davis$^{3}$}
\address{$^1$ Physics Department, Victoria University, Wellington, New Zealand}
\address{$^2$ Institut f{\"u}r Theoretische Physik,
Universit{\"a}t Innsbruck, 6020 Innsbruck, Austria}
\address{$^3$ Physics Department, University of Otago, Dunedin, New 
Zealand}
\maketitle
\begin{abstract}
The formation process of a Bose-Einstein condensate in a trap is described 
using a master equation based on quantum kinetic theory, which can be well 
approximated by a description using only the condensate mode in interaction 
with a thermalized bath of noncondensate atoms. A rate equation of the form
$ \dot n = 2W^+(n)\left\{\left(1-e^{\{\mu_n -\mu\}/kT}\right)n +1\right\}$ is 
derived, in 
which the difference between the condensate chemical potential $ \mu_n$ and 
the bath chemical potential $ \mu$ gives the essential behavior.  Solutions 
of this equation, in conjunction with the theoretical description of the 
process of evaporative cooling,  give a characteristic latency period for
condensate formation and appear to be consistent with the observed behavior of 
both rubidium and  sodium condensate formation.
\end{abstract}
\narrowtext
\pacs{PACS Nos. }
The experiments on Bose-Einstein condensation of dilute atomic gases 
\cite{JILA,MIT,RICE} have stimulated theoretical effort, which has however not 
produced any definitive result for the growth of the condensate from the 
vapor, although there have been significant theoretical contributions 
\cite{KaganKinetics,Trento,Stoof91,Semikoz,Anglin}.
This paper will present a {\em quantitative} and {\em experimentally testable} 
description of the growth process,
based on quantum kinetic theory \cite{QKI,QKII}, which can be simplified to a 
single first order differential equation for the number $ n $ of atoms in the 
condensate.  

Our formulation
contains the following principal features.  We use the Hamiltonian
\begin{eqnarray}\label{Desc1.2}
H&=&\int d^3{\bf x}\,\psi ^{\dagger }({\bf x})
\left(- {\hbar ^2\over2m}\nabla ^2\right) \psi ({\bf x})   
\nonumber \\ &&
+{1\over 2}\int d^3{\bf x}\int d^3{\bf x}'\psi ^{\dagger }({\bf x})
\psi ^{\dagger }( {\bf x}') u( {\bf x}-{\bf x}') 
\psi ( {\bf x}') \psi ( {\bf x}) 
\nonumber \\ && +
 \int d^3 {\bf x}\,V_T({\bf x)}\psi^\dagger({\bf x})\psi({\bf x})
. 
\end{eqnarray}
The potential function $ u( {\bf x}-{\bf x}')$ is as usual 
not the true interatomic potential, but rather a short range 
potential---approximately of the form $ u\delta({\bf x}-{\bf x}')$---which 
reproduces the correct scattering length.  \cite{B-approx}

We divide the condensate into two regions called the {\em condensate 
band} $R_{C}$, and the {\em noncondensate band} $R_{NC}$, as  in 
Fig.1.  We treat $ R_{NC}$ as being thermalized,
representing the majority of the atoms as a heat bath which 
provides the source of atoms for condensate growth.
The condensate band is the region of energy levels less than a value $ E_R$,
which includes not only the ground state, in which the condensate forms,
but also those levels which would be significantly affected by the presence of 
a condensate.  \cite{Justification}
In the noncondensate band, with energy levels greater than $ E_R$,
there is no significant such effect.  

The behavior in $R_{C}$ is treated fully quantum-mechanically, 
and a description in terms of trap levels modified by the presence of 
a condensate is used.  At any time there is a given 
number $N$ of atoms in $R_{C}$, and the energy levels in such a 
situation can be 
described using the number-conserving Bogoliubov 
method devised by one of us \cite{trueBog}, so that the state of 
$R_{C}$  is fully described by the {\em total number} of 
atoms $N$ in $R_{C}$, and the quantum state of the quasiparticles 
within $R_{C}$. 
In this formulation we can write the condensate band field operator in the form
\begin{eqnarray}\label{quasi}
\psi_C({\bf x)} &=& B\left\{\xi_N({\bf x}) 
+\sum_m{ b_m f_m({\bf x}) + b_m^\dagger g_m({\bf x})\over\sqrt{N}}
\right\}.
\end{eqnarray}
The quasiparticles, of energy $ \epsilon^m_N$, are 
described by annihilation operators $ b_m$,  while $ B^\dagger$ is the creation 
operator which takes the $ R_C$ system, for any $ N$, from the ground state 
with $ N$ atoms to the ground state with $ N+1 $ atoms.
The condensate wavefunction is $ \xi_N({\bf x})$, and this satisfies the 
Gross-Pitaevskii equation
\begin{eqnarray}\label{tb10}
-{\hbar^2\over 2m}\nabla^2\xi_N 
+V_T\xi_N + N u \bigl |\xi_N\bigr |^2 \xi_N &=& \mu_N \xi_N .
\end{eqnarray}
The amplitudes $ f_m({\bf x})$, $ g_m({\bf x})$ are for creation and 
destruction of quasiparticles of energy $ \epsilon^m_N$, and are fully defined 
in 
\cite{QKII,trueBog}, but will not play any significant part in this paper.

In this number-conserving Bogoliubov 
method, the {\em atoms} are conserved, while the quasiparticles 
are mixtures of {\em phonon} states, and these phonons relate to the process of 
transferring an atom from an excited quantum state to the condensate level.  
Thus, the operators $ b_m,b_m^\dagger$ do not change the total numbers of 
particles, while the operator $ B$, which multiplies everything else in 
(\ref{quasi}), reduces the total number of particles by 1.

The process we wish to describe is as follows:

(i):  Some of the collisions in $ R_{NC}$ will transfer an atom to 
 $ R_C$, so that $ N \to  N + 1$, and there is of course the reverse process 
where a collision of a noncondensate band atom with one within the 
condensate band transfers an atom from the condensate band into the 
noncondensate band, so that $ N\to N-1$.
(ii):
We consider a situation in which there is initially no condensate---however 
the boundary between $ R_C$ and $ R_{NC}$ is fixed to be appropriate for the 
amount of condensate which is {\em finally} formed.
(iii):
By evaporative cooling, the chemical potential of the atoms in $ R_{NC}$ 
becomes nonnegative; this is permissible provided 
the chemical potential does not exceed the lowest energy $ E_R$ of $ R_{NC}$.  
(iv):  With a weak interaction potential $ u $ the Bogoliubov spectrum and 
wavefunctions are valid for {\em all} $ N$, large and small, since for small 
$ N $ and $ u $ the results are not significantly different from perturbation 
theory.

Using quantum kinetic theory \cite{QKII} it is possible to derive a simple 
master equation for the density operator $ \rho$ which describes the state of 
the condensate.  The main processes are caused by 
an atom scattering into or out of
$ R_{C}$, and this can occur in six ways; that is $ N \to N\pm 1$ with one of; 
no change in the number of quasiparticles, the creation of a quasiparticle or 
the absorption of a quasiparticle.

The six transition probabilities can now all be written in terms of the 
functions $ R^\pm$
as
\begin{eqnarray}\label{new3}
W^{+}(N) &=& R^+(\xi_{ N},\mu_N/\hbar)
\\ \label{new3a}
W^{-}(N) &=& R^-(\xi_{N-1},\mu_{N-1}/\hbar)
\\ \label{new4}
W_m^{++}(N) &=& R^+\big(f_m,(\epsilon^m_N+\mu_N)/\hbar\big)
\\ \label{new5a}
W_m^{--}(N) &=& R^-\big(f_m,(\epsilon^m_{N-1}+\mu_{N-1})/\hbar\big)
\\ \label{new5}
W_m^{+-}(N) &=& R^+\big(g_m,(-\epsilon^m_N+\mu_N)/\hbar\big)
\\ \label{new4a}
W_m^{-+}(N) &=& R^-\big(g_m,(-\epsilon^m_{N-1}+\mu_{N-1})/\hbar\big)
\end{eqnarray}
The functions
$ R^{\pm}(y,\omega)$ are defined by
\widetext
\begin{eqnarray}\label{new1}
 R^{+}(y,\omega)&=&{u^2\over(2\pi)^5\hbar^2}\int d^3{\bf x}
\int d\Gamma\,\Delta(\Gamma,\omega)
F_1F_2 (1+F_3) W_y({\bf x},{\bf k})
\\ \label{new2}
 R^{-}(y,\omega)&=&{u^2\over(2\pi)^5\hbar^2}\int d^3{\bf x}
\int d\Gamma\,\Delta(\Gamma,\omega)
( 1+F_1)( 1+F_2)F_3 W_y({\bf x},{\bf k})
\end{eqnarray}
\narrowtext
where we will use the notation
\begin{eqnarray}\label{Gamma}
d\Gamma &\equiv&
 d^3{\bf K}_1 d^3{\bf K}_2 d^3{\bf K}_3 d^3{\bf k}
 \\ \label{Delta}
\Delta(\Gamma,\omega) &\equiv&
\delta(\Delta\omega_{123}({\bf x})-\omega)
\delta({\bf K}_1 + {\bf K}_2 - {\bf K}_3 - {\bf k})
.
\end{eqnarray} 
Here we use the notation 
\begin{eqnarray}\label{f10}
 W_y({\bf x},{\bf k}) = {1\over (2\pi)^3}\int d^3{\bf v}\,
y^*\left({\bf x}+{{\bf v}\over 2}\right)y\left({\bf x}-{{\bf v}\over 2}\right)
e^{i{\bf k}\cdot{\bf v}}
\end{eqnarray}
to represent the Wigner function 
corresponding to the wavefunction $ y({\bf x})$.
The function $F_i\equiv F({\bf K}_i,{\bf x})$ is the noncondensate atom  
density per $ h^3$ of phase space, and 
\begin{eqnarray}\label{f11}
\mbox{and }\quad\Delta\omega_{123}({\bf x}) &=&
 \omega_{{\bf K}_1}({\bf x})+
\omega_{{\bf K}_2}({\bf x})-
\omega_{{\bf K}_3}({\bf x})
\\
\label{f12}
\mbox{ with }\quad \hbar\omega_{{\bf K}}({\bf x})
&=& {\hbar^2{\bf K}^2\over 2m} + V_T({\bf x}).
\end{eqnarray}
We can write a stochastic master equation for the occupation probabilities
$ p(N,{\bf n})$, where $ {\bf n} =\{n_m\}$, the set of all quasiparticle 
occupation numbers, in the form \cite{offdiag}
\begin{eqnarray}\label{stochastic}
\dot p(N,{\bf n}) &=& 2NW^+(N-1)p(N-1,{\bf n})-2(N+1)W^+(N)p(N,{\bf n})
\nonumber \\  &+&
2(N+1)W^-(N+1)p(N+1,{\bf n})-2NW^-(N)p(N,{\bf n})
\nonumber \\  &+&\sum_m\{ 2n_mW_m^{++}(N-1)p(N-1,{\bf n}-{\bf e}_m)
                   -2(n_m+1)W_m^{++}(N)p(N,{\bf n})\}
\nonumber \\  &+&\sum_m\{ 2(n_m+1)W_m^{--}(N+1)p(N+1,{\bf n}+{\bf e}_m)
                   -2n_mW_m^{--}(N)p(N,{\bf n})\}
\nonumber \\  &+&\sum_m\{ 2(n_m+1)W_m^{+-}(N-1)p(N-1,{\bf n}+{\bf e}_m)
                   -2n_mW_m^{+-}(N)p(N,{\bf n})\}
\nonumber \\  &+&\sum_m\{ 2n_mW_m^{-+}(N+1)p(N+1,{\bf n}-{\bf e}_m)
                   -2(n_m+1)W_m^{-+}(N)p(N,{\bf n})\}
\end{eqnarray}
Here $ {\bf e}_m = \{\dots 0,0,1,0,0,\dots \}$ has its only nonzero value
at the position corresponding to the index $ m$. 

We can interpret functions 
$ R^\pm(y,\omega)$ as forward and backward collision rates for those collisions 
which result in a particle entering ($ +$) or leaving ($ - $) the condensate 
with an energy $ \hbar\omega$.  
The collision must 
also take place in a position where the condensate Wigner function is nonzero. 
The terms in (\ref{stochastic}) representing transitions to the ground state of 
the condensate 
exhibit a stimulated increase in collision rate of 
approximately $ N$, which 
can be a number up to $ 10^7$, 
but the transition probabilities 
$ W^{\pm\pm}_m$ defined 
in (\ref{new4}--\ref{new4a}) are multiplied only by $ n_m$, which
does not become large.
Thus as an initial approximation we drop the quasiparticle terms, which are 
smaller 
by a factor of $ N$, and are thus negligible for most of the condensation 
process.  We note that the condensate wavefunction is in practice sharply 
peaked at $ {\bf x} = {0}$ by comparison with the phase space distribution 
function $ F({\bf K},{\bf x})$ , and thus replace $ {\bf x}$ wherever it occurs 
by 0, except in $ W_{\xi_N}({\bf x},{\bf k})$, whose integral gives the 
$ {\bf k}$ 
space condensate probability density $ |\tilde\xi_N({\bf k})|^2$.  We finally 
get the simple master equation which consists of only the first line of
(\ref{stochastic}), and in which the transition matrix elements take the 
simplified form
\begin{eqnarray}\label{simp2}
W^+( N)&=& {u^2\over(2\pi)^5\hbar^2}
\int d\Gamma\Delta\left(\Gamma,{\mu( N)/\hbar}\right)
\nonumber\\&&\times
F_1F_2( 1+F_3)
|\tilde \xi_N({\bf k})|^2
\\ \label{simp3}
W^-( N)&=& {u^2\over(2\pi)^5\hbar^2}
\int d\Gamma\Delta\left(\Gamma,{\mu( N)/\hbar}\right)
\nonumber\\&&\times
( 1+F_1)( 1+F_2)F_3|\tilde \xi_N({\bf k})|^2
\end{eqnarray}
The evaluation of $ W^\pm$ can be done in various degrees of approximation; we
shall take here a thermal $ R_{NC}$ with
\begin{eqnarray}\label{simp5}
F({\bf K},{\bf x}) 
\approx\left[ e^{(\hbar\omega_{\bf k}+V_T({\bf x}) -\mu)/kT}-1 \right]^{-1}
\end{eqnarray}
from which one easily obtains (choosing $ V_T({\bf 0})=0$)
\begin{eqnarray}\label{simp6}
W^+(N) =e^{\{\mu -\mu_N\}/kT}W^-(N),
\end{eqnarray}
which indicates that a steady state is reached at large $ N$ when the chemical 
potential of the condensate almost equals that of the noncondensate. 
(Inclusion of the smaller terms in the master equation also shows that the  
temperatures of $ R_C$ and $ R_{NC}$ are equal in equilibrium).  Evaluation of 
$ W^+$ can be done by taking the energy range of 
$ R_{C}$ to be negligible compared to that of $ F({\bf K},{\bf 0})$, and by 
taking the range of $ {\bf k}$ to be small compared to that of $ {\bf K}$ in 
$ F({\bf K},{\bf 0})$.  
We also approximate the Bose function (\ref{simp5}) by its Boltzmann equivalent 
for most of the range of integration since the integrals can then be evaluated 
analytically---however this is a purely technical issue, which does not affect 
the essence of the results.
Using $ u = 4\pi a\hbar^2/m$, where $ a$ is the 
$s$-wave scattering length, we get
\begin{eqnarray}\label{simp7}
W^+(N) &=& {4m (akT)^2 \over \pi \hbar^3}e^{2\mu/kT}
\left\{{\mu_N\over kT}K_1\left({\mu_N\over kT}\right)\right\}.
\end{eqnarray}
(Here $ K_1(z)$ is a modified Bessel function. Notice also that the prefactor 
$  {4m (akT)^2 /\pi \hbar^3} $ is essentially the elastic collison rate 
$ \rho\sigma v $ , where the quantities are evaluated at the critical point for 
condesation.)
Under the assumption that the majority of the atoms are in the condensate, the 
major behavior of the master equation (\ref{stochastic}) is given by the 
rate equation for the mean number of atoms in the condensate 
(written as $ n$),
\begin{eqnarray}\label{simp8}
\dot n = 2W^+(n)\left\{\left(1-e^{\{\mu_n -\mu\}/kT}\right)n +1\right\}.
\end{eqnarray}
Since (\ref{simp6}) has been used, this represents a situation
in which  a condensate  	(which may be initially unoccupied) is in contact with 
a bath of noncondensed atoms.  	 If (\ref{simp8}) is used alone, it describes a
situation in which the thermal bath is not depleted as the condensate
evolves. A very simple form for the differential equation can be given in
this case by using (\ref{simp7}),  together with a   harmonic trap potential
$ V_T({\bf x}) = m(\omega_x^2x^2 + \omega_y^2y^2 + \omega_y^2z^2)/2 $;
and the Thomas-Fermi approximation 
$\mu_N = 
\left({15Nu\omega_x\omega_y\omega_zm^{3/2}/16\pi\sqrt{2}}\right)^{2/5},
$
(with, however, a linear interpolation as $ N\to 0$ to give the correct 
noninteracting value of $ \mu(0) = \hbar(\omega_x+\omega_y+\omega_z)/2$) 
yields a differential equation which can be 
easily integrated numerically.  

We present solutions for the parameters of the original rubidium
\cite{JILA} and sodium \cite{MIT}
experiments in Fig.2.  In both cases 
there is a latency time during which the condensate is initiated by the 
spontaneous term (the last term inside the curly brackets in (\ref{simp8}),
after which the stimulated term (the term proportional to $ n $ inside the 
curly brackets in (\ref{simp8})) takes over, causing a rapid growth until
saturation sets in when the condensate chemical potential $ \mu_n$ approaches
the chemical potential $ \mu $ of the bath. 

The timescales for the growth of the condensate are  of the same order 
of magnitude as experimentally observed, although no 
measurements have been published.  In comparing with 
experiment, one should bear in mind that this treatment (i) 
neglects fluctuation effects and (ii) neglects the quasiparticle effects.  
Inclusion of these should not change the growth curve in the region where $ n $ 
is substantially larger than 1, but could possibly speed up the process by 
which the first 100 or so atoms enter the condensate. 

The solutions in Fig.2 assume that $ \mu$ and $ T$, the chemical 
potential and temperature of the bath of noncondensed atoms, are constant.  
They 
nevertheless exhibit the fundamental nature of the process of condensation.  
For a treatment more appropriate to comparison with experiment one must couple 
the condensate growth equation (\ref{simp8}) to appropriate time development 
equations for the process of evaporative cooling, such as those of 
\cite{Holland,Luiten,Davis}.  There are three principal timescales in the 
problem; the timescale of equilibration of the noncondensate ``bath'', which 
is very fast, the timescale of condensate growth, as given by solutions of 
(\ref{simp8}), and the timescale of the evaporative cooling process, which is 
in practice considerably slower than both of the others.  Under these 
conditions one would expect that a model in which the noncondensate ``bath'' 
distribution function is considered to be always thermalized for particles 
below the ``cut'' energy, which we shall call $ \eta k T $, and is zero 
above 
this ``cut'' energy would be valid.  Provided the cooling process is 
slow enough, we can use (\ref{simp6}) with the resulting time-dependent 
$ \tilde T$ and $ \tilde \mu$ (values appropriate 
to the truncated distribution), after modifying the evaporative cooling
equations to take account of the transfer of particles and energy between 
condensate and noncondensate.

One should also note that the Boltzmann function with a 
cutoff at the energy $ \eta kT$, typically 
with $\eta = 5$--$7$, is a distribution which is significantly out of 
thermal equilibrium; for $ \eta= 5$ or $7 $ we find respectively 12.5\% and
 2.9\% of the full Boltzmann 
distribution is above $ \eta kT$.
However collisions which are of the correct kinematic configuration to 
permit one of the atoms to enter the condensate are a selection of the full 
Boltzmann distribution in which the density of states factor, proportional to 
$ E^2$ for the harmonic oscillator, becomes approximately independent of $ E$; 
for them only 0.67\% and 0.091\% have energy 
greater than $ \eta kT$.
Thus the truncated distribution behaves like a genuine 
thermal distribution for the collisions which can populate the condensate.

	 Using this more complete model of condensate formation, we have
	 simulated a number of different evaporative cooling paths that are
	 consistent with the published descriptions of Bose condensate formation
	 (e.g \cite{JILA,MIT}). Although differing in detail, the results are broadly 
in agreement with those shown in Fig 2. One of the features of the
experimental process is that the cooling process is halted, and the
system allowed to thermalize for a short period before the condensate
is observed. Our simulations show that the nonequilibrium noncondensate
distribution evolves to a true Boltzmann distribution in  a few mean
collision times, and it is easy to show that the new temperature is
always less than $ \tilde T$, while the new chemical potential may be
larger or smaller than $ \tilde \mu$. Typically, in the regime appropriate
to the experiments,  $  \mu$ increases 	during this relaxation, and	 may
change from a value below to a value above $\mu_C $ . This crucial  step
in the formation of the condensate, of $ \mu$ evolving to exceed $\mu_C $,
may thus have occurred only during the  relaxation 	process.

The condensate growth equation (\ref{simp8}) is  like the kind of equation 
one finds for laser.  Thus there is the spontaneous emission term (the $ +1$ 
inside the curly brackets), and a gain term.  The gain here is determined 
entirely by the difference between the condensate chemical potential 
$ \mu_n$, which is quantum mechanically determined, and that of the 
noncondensate ``bath'' $ \mu$, which is determined by statistical mechanics.  
The fact that the quasiparticles play no significant role in the process
is analogous to the behavior of a multimode laser, in which nearly all
photons go into the mode with the highest gain, even if it is only marginally 
the highest gain.

One can also conclude that adaptations of the condensate growth equation for 
different configurations of the noncondensate ``bath'' will prove a useful 
tool in the eventual design of an atom laser, or ``Boser''.

This work was supported by the Marsden Fund under contract number
PVT-603, and by \"Osterreichische Fonds zur F\"orderung der 
wissenschaftlichen Forschung.

\begin{figure}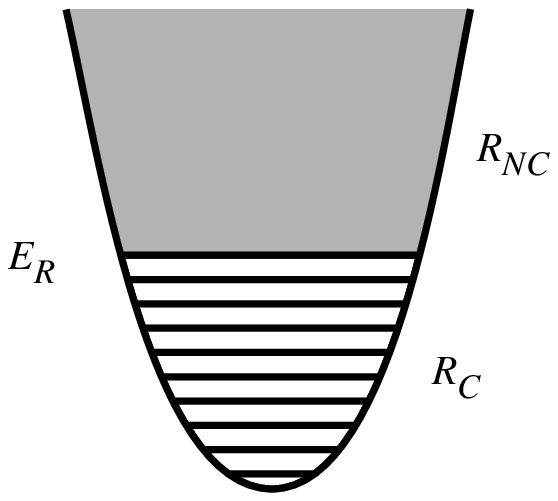
\label{fig1.eps}
\caption{Fig.1. The condensate and noncondensate bands}
\end{figure}

\begin{figure}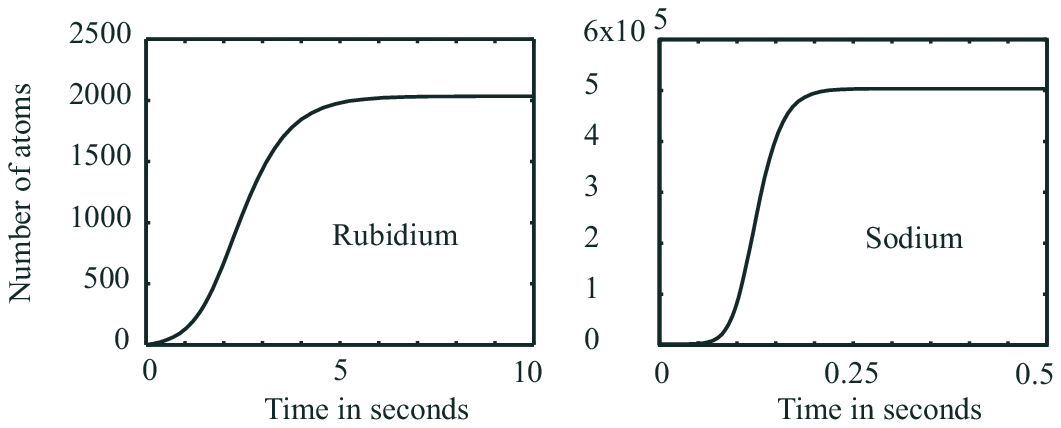
\label{fig2.eps}
\caption{Fig.2. Condensate growth for (a) Rubidium 
and (b) Sodium. Scattering lengths are $ 5.71\,{\rm nm}$ and 
$ 2.75\,{\rm nm}$, respectively.}
\end{figure}%


\begin{references}
\bibitem{offdiag} A full master equation which treats off-diagonal elements 
can be similarly derived and will be presented elsewhere \cite{QKII}.
\bibitem{Justification}Numerical investigations show that for the usual trap 
potentials the choice $ E_C \approx 3\mu_N$ 
means that the higher levels are equal to those of the noncondensed system to 
at least an accuracy of  5\%.  Typically, $ \mu_N$ is about an order of 
magnitude less than $ kT$. Under these conditions a thermal distribution of 
energies corresponds to over 95\% of the atoms being thermalized and 
within $ R_{NC}$.  Even after condensation, the majority of the atoms are still
in $ R_{NC}$.
\end{references}
\end{document}